%% file: template.tex
\documentclass{article}

\usepackage{arxiv}

\usepackage[utf8]{inputenc} 
\usepackage[T1]{fontenc}    
\usepackage{hyperref}       
\usepackage{url}            
\usepackage{booktabs}       
\usepackage{amsfonts}       
\usepackage{nicefrac}       
\usepackage{microtype}      
\usepackage{lipsum}
\usepackage{graphicx}
\graphicspath{ {./figures/} }

\usepackage{graphicx}
\usepackage{amsmath}
\usepackage{calc}
\usepackage{float}
\usepackage{subfig}
\usepackage{amsfonts}
\usepackage[usenames,dvipsnames,svgnames,table]{xcolor}
\usepackage{amsthm}
\usepackage[utf8]{inputenc}
\usepackage{listings}
\usepackage{xcolor}
\usepackage{appendix}

\title{Physics Informed Deep Learning for flow and transport in porous media}

\author{
  Cedric G. Fraces \\
  Department of Energy Resources Engineering\\
  Stanford University\\
  Stanford, CA 94305 \\
  \texttt{cfraces@stanford.edu} \\
   \And
 Hamdi Tchelepi \\
  Department of Energy Resources Engineering\\
  Stanford University\\
  Stanford, CA 94305 \\
  \texttt{tchelepi@stanford.edu} \\
}

\begin{document}
\maketitle
\begin{abstract}
We present our progress on the application of physics informed deep learning to reservoir simulation problems. The model is a neural network that is jointly trained to respect governing physical laws and match boundary conditions. The methodology is hereby used to simulate a 2-phase immiscible transport problem (Buckley-Leverett). The model is able to produce an accurate physical solution both in terms of shock and rarefaction and honors the governing partial differential equation along with initial and boundary conditions. We test various hypothesis (uniform and non-uniform initial conditions) and show that with the proper implementation of physical constraints, a robust solution can be trained within a reasonable amount of time and iterations. We revisit some of the limitations presented in previous work \cite{Fuks2020} and further the applicability of this method in a forward, pure hyperbolic setup. We also share some practical findings on the application of physics informed neural networks (PINN). We review various network architectures presented in the literature and show tips that helped improve their convergence and accuracy. The proposed methodology is a simple and elegant way to instill physical knowledge to machine-learning algorithms. This alleviates the two most significant shortcomings of machine-learning algorithms: the requirement for large datasets and the reliability of extrapolation. The principles presented can be generalized in innumerable ways in the future and should lead to a new class of algorithms to solve both forward and inverse physical problems.
\end{abstract}

\input{BL.tex}
\pagebreak

\bibliographystyle{unsrt}  
\bibliography{references}

\end{document}

%% file: BL.tex
\section{Introduction}
Earth systems predictions largely rely upon the modeling of complex fluid dynamics in the porous media characterizing the subsurface. In order to support better decision making, engineers and geo-scientists have developed forecasting tools and algorithms that are based on various delineations of mass and energy conservation along with the empirical Darcy law characterizing flow in porous media. Models must combine physical principles and observations in order to make accurate prediction. It is therefore important to quantify these two features:
\begin{enumerate}
    \item The complexity of the model in terms of scales and physics they emulate
    \item Their ability to assimilate data
\end{enumerate}
Numerical reservoir simulation lies on one end of that (complexity-calibration) spectrum as they can capture complex physics and multi-scale effects (heterogeneities, multi-phase, compositional, thermal, chemical reactions, geomechanics,...). The resolution of large numerical systems involve time intensive computations and it is not rare to encounter models with single run-time of several hours to days. On the other hand, decline curve analysis (DCA, \cite{Arps1945}) is based on very simple physics that can easily be calibrated to observed data. Because of the large degree of uncertainty characterizing energy systems, most predictive models are parametric and must be updated against observed data. This gives rise to two classes of problems:
\begin{enumerate}
    \item The forward (or inference) problem where the goal is to make prediction on the time evolution of a quantity of interest (QOI) -typically the production of a field- based on a trusted model
    \item The inverse (or identification) problem where the goal is to estimate the parameters of the model -the subsurface properties for example- that honor the observation on the QOI and a prior understanding we have of the system.
\end{enumerate}
 
Energy systems present a unique set of challenges to machine learning methods. They have large financial and environmental footprints, wide uncertainties and obey strict physical constraints. Recent advances in instrumentation, telemetry and data storage have allowed operators to rely more and more on data to make decisions. However, the integration of all this data represents a challenge. While the need to inform decisions in a timely manner has become an essential competitive differentiator in most industries, \cite{DeBezenac2017}. One of the main motivation for the present work is that to this day, we lack a robust methodology for the application of machine learning to Earth systems, \cite{Reichstein2019}. The challenges identified are:
\begin{itemize}
    \item Data assimilation
    \item Scale and Resolution
    \item Generalization outside training scope
    \item Interpretability and causal discovery
    \item Physical consistency of governing physical laws
    \item Uncertainty Quantification
    \item Limited access to labeled data
\end{itemize}
In this work, we propose to address some of these issues. We will be leveraging some of the most recent advances in statistical and deep learning and more particularly the use of deep neural networks. We plan to utilize some of their known advantages: their ability to model complex data structures all while limiting some of their shortcomings: their inability to generalize well and to provide interpretable solutions. After a literature review of existing state of the art  physics-informed machine learning techniques, we present our method and its application to the Buckley-Leverett problem \cite{BuckleyLeverett1942}. We extend with variations of the original problem including multiple physics and dimensions.

\section{Related work}
\label{sec:related_work}
\subsection{Data driven approaches}
In spite of a large interest and commercial needs, there is relatively little applied research on data driven solutions for reservoir engineering. Classical engineering tools with a proven track record (DCA, type curves, material balance, Capacitance and Resistance models, Analytical methods, numerical simulation) are being used by a vast majority of engineers and geo-scientists. Attempts to leverage data and statistical approaches are relatively new and coincide with the rise of unconventional production in the US. These are characterized by a large increase in the number of wells drilled in a short time period combined with a limited understanding of the physics governing the behavior of shale formations. Due to the availability of vast amount of data, researchers and engineers have started experimenting with data-driven approaches to get insights into the production patterns. The present work was initiated with a method to generalize the concept of decline to cases where it normally fails (unconventional, secondary recovery, infill drilling). Mohaghegh et al.\ \cite{Mohaghegh2011} developed a top-down approach for modeling and history matching of shale production based on statistical and pattern recognition models and applied their approach to three shale reservoirs. Sun et al.\ \cite{Sun2018} applied Long-Short Term Memory algorithm (LSTM) to predict a well's oil, water, and gas production. Their work led to an improvement of the production forecast in comparison with standard DCA models. However, the tested portion of their time series exhibits weak variations in comparison with the training part and it remains unclear whether the algorithm can generalize well to complex time series. 

\subsection{Physics Informed}
Others have applied Physics Informed Neural Networks to the problem of single well production forecast. These results encouraged the expansion to the more complex problem of 2-Phase transport in one dimension and then in more dimensions. Chiramonte (\cite{Chiarmaonte2018}) shows two examples of resolution of the Laplace equation in two dimensions. They present some error analysis and compare the solution with traditional finite volume and show that the solution obtained using neural networks loses some resolution.
Rudy (\cite{Rudy2018}) presents a study on the data driven identification of parametric partial differential equations. They use multivariate sparse regression techniques (Lasso, Ridge) to calculate the coefficients of Burger and Navier-Stokes equations. Their approach allow the coefficients to have arbitrary time series, or spatial dependencies. Considering the large collection of candidate terms for constructing the partial differential equation, it is assumed that a sufficiently large library of 'base' functions is used to approximate the PDE's coefficient. The approach, although very promising suffers two major drawbacks: first, the use of numerical differentiation in their gradient descent optimization; second, the assumption of sparse representation and multivariate regression is valid only if the library of basis function is sufficiently rich. This reveals untrue or impractical for most cases. Zabaras, et al.\ (\cite{MoZabaras2018}, \cite{YinhaoZabaras2018} and \cite{YinhaoZabaras2019}) use a convolutional encoder decoder approach to build probabilistic surrogates for multiple gaussian realizations of a permeability field. Their models are enriched by a physics based loss computed using Sobol filters to mimic the differential forms of a PDE. This approach is similar to the one presented by Bin Dong et al.\ (\cite{Long2018_pdenet}) .Their approach is fast and treats the problem as one of image to image regressions. If the method shows promises for elliptic equations (flow), they lack the ability to extrapolate well especially in the hyperbolic transport problem. We draw inspiration from the work of Raissi and Perdikaris, \cite{RaissiJML2018}, \cite{raissi2017physicsI}, \cite{raissi2017physicsII} who present a new method that addresses the two issues aforementioned by assigning prior distributions in the form of artificial neural networks or Gaussian processes. Derivatives of the prior can now be evaluated at machine precision using symbolic or automatic differentiation (Deep learning libraries such as Tensorflow, \cite{tensorflow2015-whitepaper} are well suited for this type of calculation). This allows a certain noise level in the observations and removes the need to manually compute the derivatives of the solution in order to evaluate the residual of the PDE. Although results are demonstrated for the Burger's equation in 1D with a sinusoidal initial condition, we show that it fails in the case of Buckley Leverett with a constant initial and boundary condition.

\subsection{A word on Automatic differentiation}
Automatic differentiation (AD) lies in between symbolic and numerical differentiation. If the method does provide values of the derivatives and keeps track of the derivative values of expressions (as opposed to their expression), it does so by establishing first symbolic rules for the differentiation. AD relies on the postulate that all derivation operations can be decomposed in a finite set of basic operations for which derivatives are known (example: $\frac{dx^2}{dx}=2x$). These basic operations can be combined through the so-called chain rule in order to recover the derivative of the original expression. This hybrid approach gets derivatives at machine precision in a time that is comparable to getting them manually (with a small overhead). \cite{Baydin2018_AD} provides a revue of AD methods and their application to machine learning. Open source software libraries for AD such as Tensorflow (\cite{tensorflow2015-whitepaper}) allow the creation and compilation of computational graphs that are then used to differentiate quantities of interest at observation points. Once computed, the graph is used through the entire optimization process (whether for training a neural network or for computing the residual of a PDE). This static implementation results in significant savings in terms of computational time.

\subsection{Network parameters and architectures}
In order to better understand the training process and provide some guidelines on methodology, we have reviewed a series of deep learning architectures recently proposed for the resolution of PDEs. We test and benchmark these formulations in a later section.

\subsubsection{SiRen}
Sitzmann et. al (\cite{SiRen2018}) proposes to use sinusoidal activation functions in the resolution of partial differential equations. 
\begin{equation}
    \phi_i(x) = sin(\mathbf{W}_ix_i+b_i)
\end{equation}
Where $i$ represents the layer of the network, $\mathbf{W}$ the weights and $b$ the biases.
The justification is that with proper distribution control, the sine function and is derivatives remain bounded in a deep architecture and lead to faster convergence on benchmark image and signal processing problems. They also help capture the high frequency signal defined implicitly as the solution to certain classes of partial differential equations. Examples are demonstrated with the in-homogeneous Helmholtz equation and the two-dimensional wave equation

\subsubsection{Adaptive Activation}
Adaptive activation functions (\cite{AdaptiveActivation2019}) introduce a scalable hyperparameter in the activation function that changes dynamically the topology of the loss function. A trainable parameter is multiplied by the input to the activation in order to modify its slope. The transformation occurring at one layer takes the form:
\begin{equation}
    \phi_i(x) = \sigma(a(\mathbf{W}_ix_i+b_i))
\end{equation}
Where $\sigma$ is a non linear activation and $a$ is the trainable parameter that effectively changes the slope of the activation function. As an example, the sigmoid function $1/1+e^{-x}$ under such transformation becomes $1/1+e^{-ax}$. The training process consists of finding the minimum of a loss function by optimizing $a$ along with the weights and biases.
The reported effect is an improvement in convergence rate, especially at early training, as well as a better solution accuracy. This idea can be complemented with connections skipping in a similar fashion as the Residual Network architecture (\cite{Long2018_pdenet}).

\subsubsection{Deep Galerkin Method}
The Deep Galerkin Method (DGM) was presented as an efficient method to compute numerical solutions of PDE's without a mesh (\cite{DGM2018}). It features an approximation for high order differentials that speeds up the computation of the PDE's residual. It consists of a series of recurring blocks composed of fully connected layers. The model $f(x, t;\theta)$ takes the following form:
\begin{equation}
\begin{aligned}
    f(x, t;\theta) &= \mathbf{W}S^{L+1} + b,\\
    S^{l+1} &= (1 - G^l)\circ H^l + Z^l\circ S^l, & l=1,\dots, L,\\
    H^l &= \sigma(U^{h,l}\vec{x} + \mathbf{W}^{h,l}(S^l\circ R^l) + b^{h,l}), & l=1,\dots, L,\\
    R^l &= \sigma(U^{r,l}\vec{x} + \mathbf{W}^{r,l}S^l + b^{r,l}), & l=1,\dots, L,\\
    G^l &= \sigma(U^{g,l}\vec{x} + \mathbf{W}^{g,l}S^l + b^{g,l}), & l=1,\dots, L,\\
    Z^l &= \sigma(U^{z,l}\vec{x} + \mathbf{W}^{z,l}S^l + b^{z,l}), & l=1,\dots, L,\\
    S^1 &= \sigma(\mathbf{W}^1\vec{x} + b^1)
\end{aligned}
\end{equation}
Where $\vec{x} = (x,t)$, $L$ the number of hidden layers and $\circ$ is the Hadamard product. The parameters of the network $\theta = \{\mathbf{W}^i, b^i\}_{\forall i}$ are the weights and biases in the fully connected layers. This architecture can be used in conjunction with a Fourier Network (\cite{Tancik2020}) where the input variable $(x,t)$ are replaced with their Fourier transforms
\begin{equation}
    \Phi_f = \{cos(2\pi ix), sin(2\pi ix), cos(2\pi it), sin(2\pi it)\}_{i\in\mathrm{N}}
\end{equation}
This architecture was proposed to help alleviate the "spectral bias" or tendency of neural networks to favor low-frequency solutions (\cite{Rahaman2019}).

\subsubsection{Fourier Neural Operator}
\cite{FourierNet2020} proposes a Neural operator based on fast Fourier transforms (FFT) that learns the mapping between a family of PDE and their parameterized solutions. The method relies on applying a Fourier transform  and updating it in an iterative fashion using: 
\begin{itemize}
    \item a Fourier transform $\mathcal{F}$ to the functional of interest $v$
    \item a linear transform $\mathcal{R}$ on the lower Fourier modes
    \item A filtering of the higher modes using the FFT
    \item An inverse Fourier transform $\mathcal{F}^{-1}$
    \item An iterative update of $v$ using a non linear transform $\sigma(Wv + \mathcal{F}^{-1}(\mathcal{F}(v))$
\end{itemize}

\section{Methodology}
The method we present uses neural networks to solve partial differential equations (PDE) of the form:
\begin{equation}
    \mathcal{R}(\mathbf{X}, t, \mathbf{\nu}, \mathbf{\nu}_t, \nabla\mathbf{\nu}, \nabla^2\mathbf{\nu},\dots) = 0
\end{equation}
In flow and transport problems, $\mathbf{\nu}$ typically refers to state variables such as the reservoir pressure and saturation $\mathbf{\nu}=[p, S]^T$.
Similarly to what we do in classical numerical analysis, we assume a representation for the unknown solution $\nu$. Finite elements formulations assume that $\nu$ is represented by a linear combination of basis functions. 
\begin{equation}
    \mathbf{\nu} \approx \sum\limits_{i} \hat{\nu}_i\phi_i
\end{equation}
In this approach instead, we assume that it is represented by function compositions with series of linear and non linear transformations. This can conveniently be represented by a feed forward multi-layer perceptron.
\begin{equation}
    \mathbf{\nu} \approx \mathbf{\nu}_{\theta} = \sigma\left[\mathbf{W}^{[n]}\times \sigma(\mathbf{W}^{[n-1]}(\dots\sigma(\mathbf{W}^{[0]}[\mathbf{X},t]^T + \mathbf{b}^{[0]}))\dots + \mathbf{b}^{[n-1]}) + \mathbf{b}^{[n]}\right]
\end{equation}
Where $\sigma$ is a nonlinear activation function (sigmoid, tanh, ReLu,...), $\mathbf{W}_i$ are weight matrices and $\mathbf{b_i}$ are bias vectors for layer $i$. As an example, a network built to emulate the unsteady water saturation in one dimension ($S_w(x,t)$) with two hidden layers featuring $N_1$ and $N_2$ nodes, a tanh activation function can be explicitly formulated as:
\begin{equation}
    S_{w,\theta}(x,t) = tanh\left( \mathbf{W}^{[2]} \left[ tanh \left( \mathbf{W}^{[1]} \left[ tanh \left( \mathbf{W}^{[0]} [x,t]^T + \mathbf{b}^{[0]} \right) \right] + \mathbf{b}^{[1]} \right) \right] + \mathbf{b}^{[2]} \right)
    \label{eq:MLP_2layers}
\end{equation}
Where $\mathbf{W}^{[0]}$ is a matrix with dimensions $[N_1,2]$, $\mathbf{W}^{[1]}$ is a matrix with dimensions $[N_2, N_1]$ and $\mathbf{W}^{[2]}$ is a matrix with dimensions $[1,N_2]$ while $\mathbf{b}^{[0]}$ is a vector with $N_1$ elements, $\mathbf{b}^{[1]}$ is a vector with $N_2$ elements and $\mathbf{b}^{[2]}$ is a scalar.

$S_{w,\theta}(x,t)$ is a continuous representation of the state variables. For any $x,t$ it will produce an output. It is also a differentiable representation. This means that we can compute any derivatives of it with respect to space and time. We can therefore construct the new quantity of interest $\mathcal{R}$ (for residual). $\mathcal{R}$ is a neural network that has the same architecture as $\mathbf{\nu}_{\theta}$, the same weights and bias, but different activation functions due to the differential operator. Several technical challenges need to be addressed in order to implement this idea. One of them is the difficulty to differentiate a functional form similar to thy defined in eq. \ref{eq:MLP_2layers} (more complex in reality) in a quick and efficient manner. We leverage the capabilities of modern software libraries like Tensorflow (\cite{tensorflow2015-whitepaper}) to compute the differential of this complex form automatically (applying the same methods used for back-propagation).

\section{Resolution of forward Riemann problem using PINN}
\begin{figure}[H]
    \centering
    \includegraphics[width=0.8\linewidth]{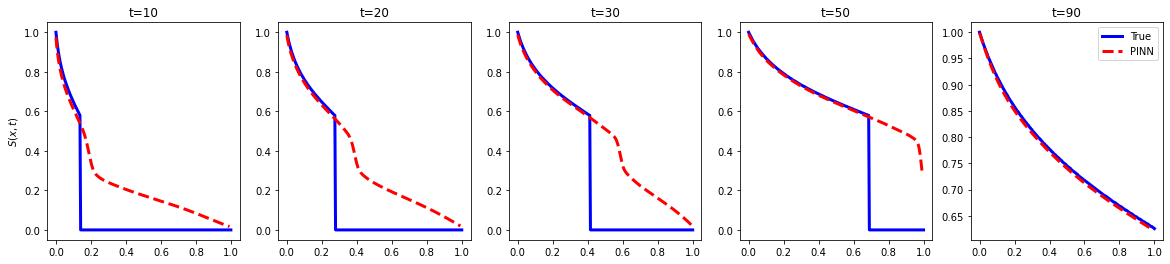}
    \caption{Results of saturation inference computed using the PINN approach (dashed red) conditioned on the weak form of the hyperbolic problem. The reference solution using MOC is plotted in blue}
    \label{fig:BL_saturation_fail}
\end{figure}

Various attempts were made at resolving the forward Buckley Leverett problem using PINNs (cf. section \ref{sec:related_work}). The straightforward application of the approach presented in \cite{raissi2017physicsI} to the hyperbolic problem with complex flux function and mixed wave solution (Eq.~\ref{eq:residual_buckley} and \ref{eq:frac_flow}) had not been conclusive so far. Fuks et. al (\cite{Fuks2020}) present a comparative study of physics informed Machine Learning (PIML) applied to non-linear one dimensional hyperbolic problems. It compares various flux functions and shows that the Riemann problem (non linear, non convex flux with multiple inflection points) cannot be solved using the presented implementation. We list some of the main characteristics of the PIML model used:
\begin{itemize}
    \item fully connected network with 8 layers and 20 neurons per layer
    \item $tanh$ activation functions
    \item loss function composed of initial and boundary conditions mean squared error (MSE) regularized with squared residual evaluated at interior collocation points
    \item LBFGS optimizer
\end{itemize}
The work concludes that PIML approximation fails to recover the accurate solution when shocks are present and is not suited for the hyperbolic PDEs with discontinuous solutions. An example of saturation front resulting from such resolution is presented in figure~\ref{fig:BL_saturation_fail}.
This conforms with the results presented in \cite{Fuks2020} and represents a baseline for the following work.

We remind that the Buckley-Leverett problem characterizing the transport of a two phase system in porous media can be solved in one dimension using various methods. The original publication by Buckley and Leverett (\cite{BuckleyLeverett1942}) shows how the mathematical integration of equation:
\begin{equation}
\label{eq:residual_buckley}
    \frac{\partial S}{\partial t} + \frac{\partial f(S)}{\partial x} =0
\end{equation}
Where the fractional flow $f$ is a nonlinear equation defined as:
\begin{equation}
    \label{eq:frac_flow}
    f(S) = \frac{(S - S_{wc})^2}{(S - S_{wc})^2 + (1 - S - S_{gr})^2/M}
\end{equation}
subject to constant boundary and initial conditions:
\begin{eqnarray}
    S(x=0,t) &= S_{inj}\\
    S(x,t=0) &= S_{wc}
    \label{eq:BL_uniform_bc}
\end{eqnarray}
Where $S_{wc}$ represents the residual (connate) saturation of the wetting phase (typically water), $S_{gr}$ the residual saturation of the non wetting phase and $M$ the end point mobility ratio between the two phases defined as the ratio of end point relative permeability and viscosity of both phases.

\begin{figure}[H]
    \centering
    \includegraphics[width=0.8\linewidth]{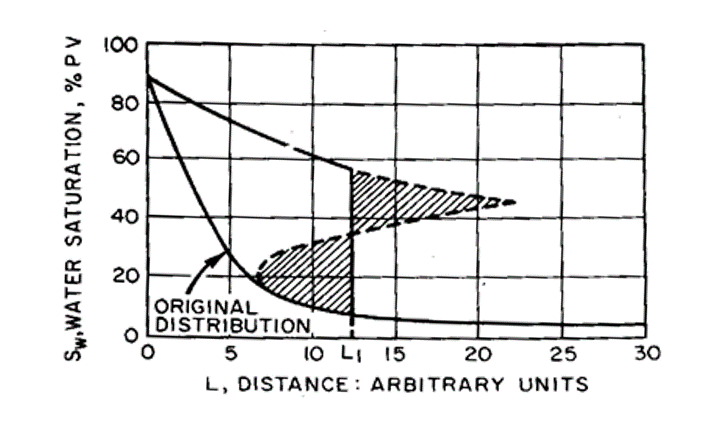}
    \caption{Triple-Valued Saturation Distribution (Source Buckley and Leverett, 1942 \cite{BuckleyLeverett1942})}
    \label{fig:BL_triple_saturation}
\end{figure}

\begin{figure}%
    \centering
    \includegraphics[width=0.5\linewidth]{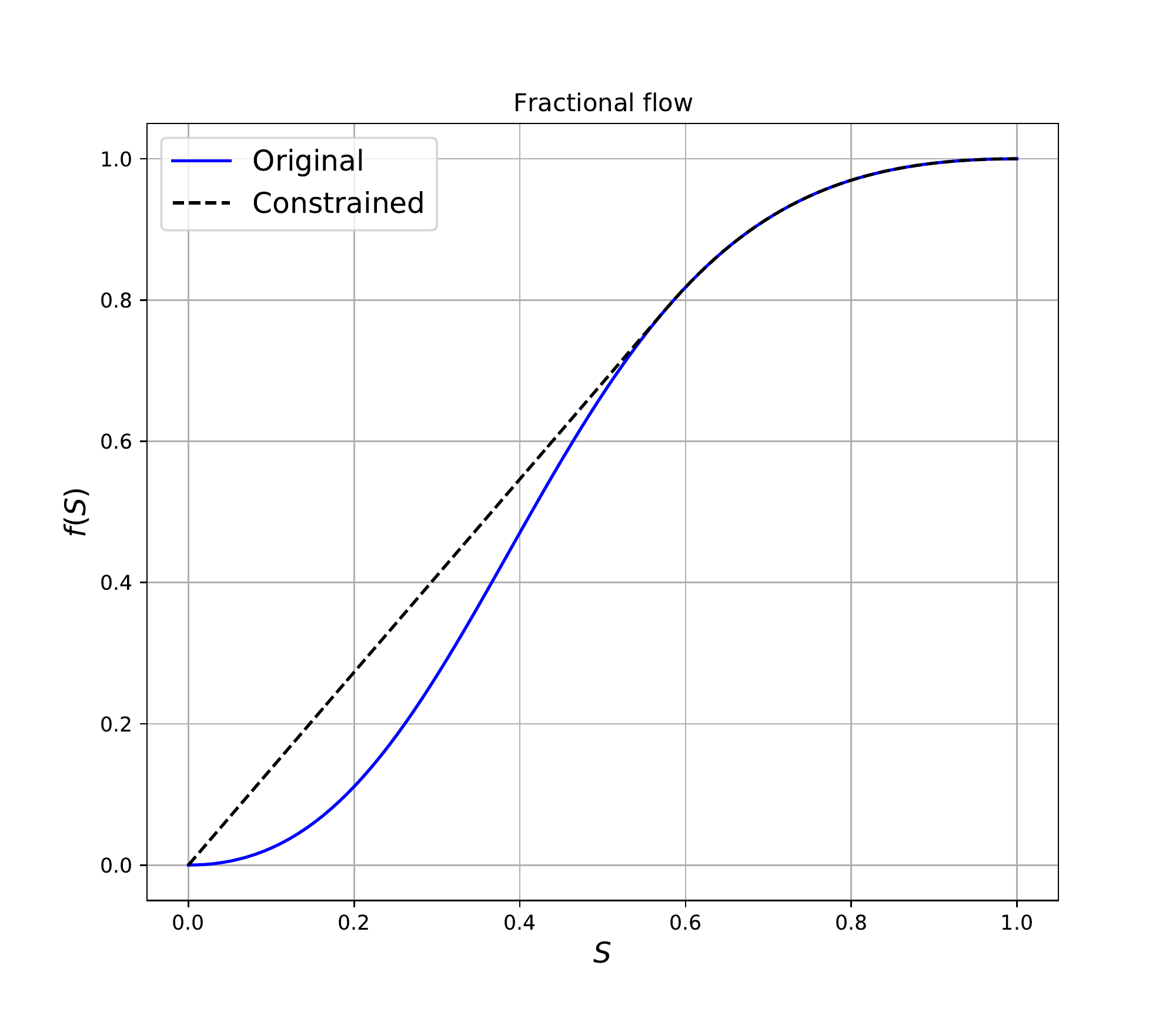}%
    \caption{Fractional flow curve (blue) along with Welge construction (dotted black) for case with $S_{wc}=S_{or}=0$}%
    \label{fig:frac_flow_welge}%
\end{figure}

The resolution of the partial differential equation \ref{eq:residual_buckley} in its weak form along with classical Dirichlet initial and boundary conditions leads to a non-physical solution. Indeed, the saturation obtained can be doubled (or tripled) valued at a given $x$ and $t$. This inconsistency is resolved  by dropping perpendiculars (shock) at flood front position so that the areas to the right equal the areas to the left of the shock as shown in Figure~\ref{fig:BL_triple_saturation}. In other words a discontinuity in $S_w$ at a flood front location $L_1$ is needed to make the water saturation distribution single valued and to provide a consistent material balance for the displacing fluid. 
Various methods have since then been proposed (\cite{welge1952}, \cite{buckley1942mechanism}) to resolve the apparent inconsistency between the weak form of the PDE described in Eq.~\ref{eq:residual_buckley} and conservation law that as an integral law may be satisfied by functions that are non-differentiable. Lax (\cite{Lax_Hyperbolic}) provides a detailed theory of conservation law featuring shock waves. This framework gives rise to the \emph{principle of entropy} that stipulates the entropy of particles crossing a shock front must increase. 
Tchelepi, \cite{ERE221} provides a detailed demonstration of how to obtain the analytical solution using the Method of Characteristics (MOC). Orr, \cite{Orr225} develops a general theory of multi-component, multi-phase displacement based on the MOC. The Buckley Leverett solution can be seen as a special case with two phases and one component per phase.
The theory shows that the differential form of the PDE along with necessary boundary and initial conditions are not enough to get a physically consistent solution to the problem. Two physical principles constraining the wave velocities are introduced in order to address the problem:
\begin{enumerate}
    \item The \emph{Velocity constraint}: Wave velocities in a two phase displacement must decrease monotonically as the upstream saturation continuously increase.
    \item The \emph{Entropy condition}: Wave velocities on the downstream side of the shock must be less or equal to the shock velocities and the velocities on the upstream of the shock.
\end{enumerate}
Figure~\ref{fig:frac_flow_welge} shows the Welge construction that encodes these two physical constraints. The displacement occurs on the convex hull of the fractional flow.
We postulate that the PINN approach must be bound by the same set of physical constraints in order to produce a physically consistent solution. Therefore we must replace the original fractional flow equation with one describing the convex hull. The parameterization is defined as:
\[
  \tilde{f}(S) =
  \begin{cases}
                                   0 & \text{if $S\leq S_{wc}$} \\
                                   \frac{S - S_{wc}}{f(S_f)} & \text{if $S_{wc}\leq S\leq S_f$} \\
  \frac{(S - S_{wc})^2}{(S - S_{wc})^2 + (1 - S - S_{or})^2/\tilde{M}} & \text{if $S>S_f$}
  \label{eq:frac_flow_welge}
  \end{cases}
\]
This piecewise form along with the underlying fractional flow are represented in figure~\ref{fig:frac_flow_welge}

\begin{figure}%
    \centering
    \includegraphics[width=1\linewidth]{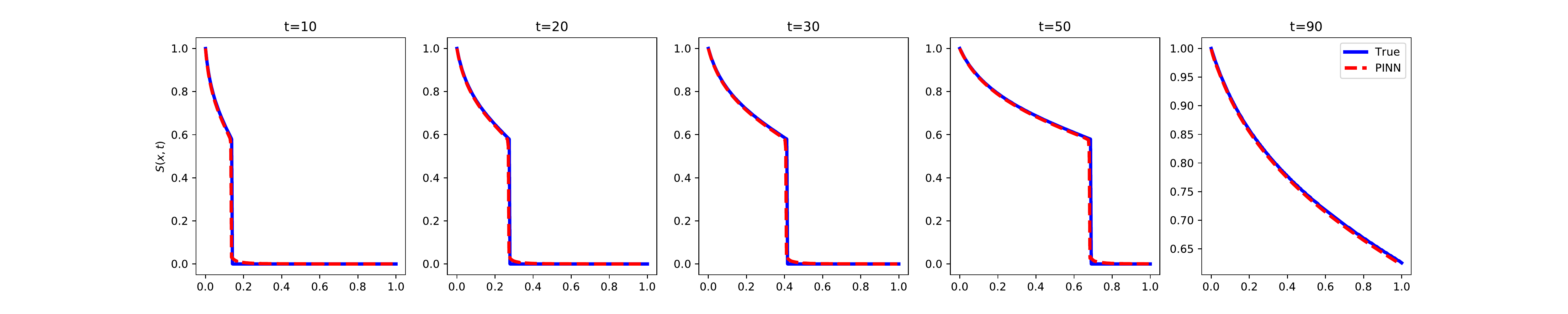}%
    \caption{Results of saturation inference using PINN (dashed red) vs MOC (blue) with velocity constraint and entropy condition. The convex hull of the fractional flow curve is used to model the displacement.}%
    \label{fig:BL_Simnet}%
\end{figure}

The next step is to differentiate the piecewise function $\tilde{f}(S)$ with respect to $x$ in order to compute the residual (Eq.~\ref{eq:residual_buckley}). This is done using the SimNet package (\cite{SimNet_guide}) which provides an interface with symbolic calculation. The results of the simulation are presented in figure~\ref{fig:BL_Simnet}

We see that the match is very close and that the shock is well captured. The $L_2$ error after $30,000$ iterations is approximately $2\times 10^{-3}$. The progression of the three losses (boundary, initial, collocation) is represented in figure~\ref{fig:BL_Simnet_loss}:

\begin{figure}%
    \centering
    \includegraphics[width=1\linewidth]{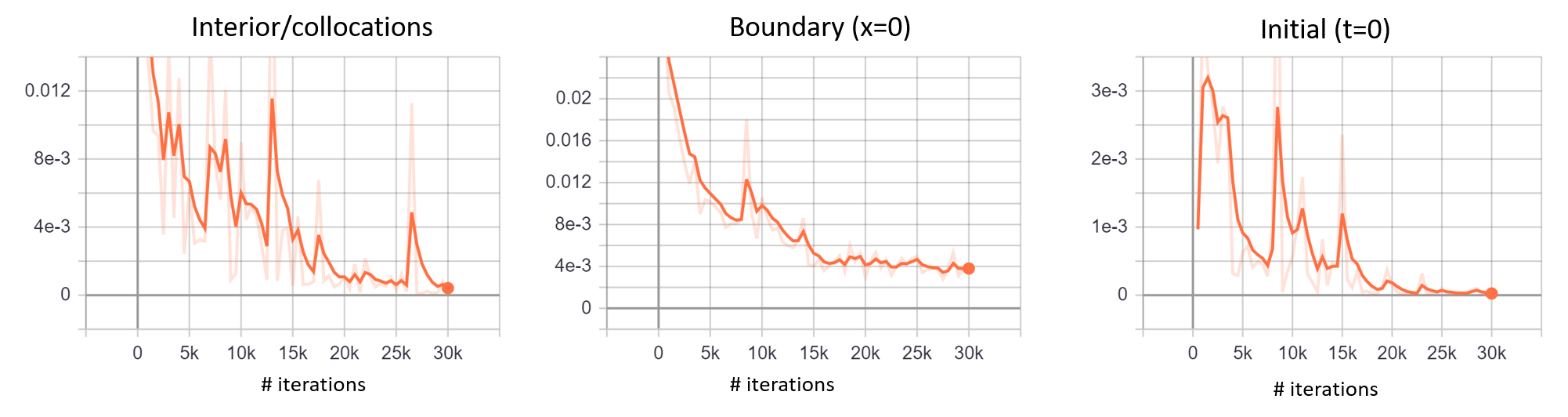}%
    \caption{Evolution of $L_2$-loss with number of iteration for: collocation points (left), boundary condition (x=0) and initial condition (t=0)}%
    \label{fig:BL_Simnet_loss}%
\end{figure}

The use of the SimNet package allows a more exhaustive testing of various architecture and hyper parameters. This combined with the introduction of physical constraints led to the resolution of the Riemann problem using PINN.

Table~\ref{table:hyperparameters} shows the nature of the parameters that were selected along with extent of the search space.
\begin{table}
\begin{center}
 \begin{tabular}{||r c||} 
 \hline
 Parameter & Values\\ [0.5ex] 
 \hline\hline
 Activation functions & 'sin', 'tanh', 'selu', 'poly', 'elu', 'swish' \\ 
 \hline
 Layer Size & 64, 256, 512, 1024 \\
 \hline
 Number of Layers & 6, 8, 12, 20 \\
 \hline
 Skip Connections & False, True \\
 \hline
 Adaptive Activation & False, True \\
 \hline
  Deep Galerkin Method & False, True\\ [0.5ex] 
 \hline
\end{tabular}
\caption{Network architectures hyper parameters selected for training on 1D Buckley Leverett problem}
\label{table:hyperparameters}
\end{center}
\end{table}

We plot the evolution of the loss functions for all the 164 experiments run and overlay the loss corresponding to fig~\ref{fig:BL_Simnet_loss} for reference.

\begin{figure}%
    \centering
    \includegraphics[width=1\linewidth]{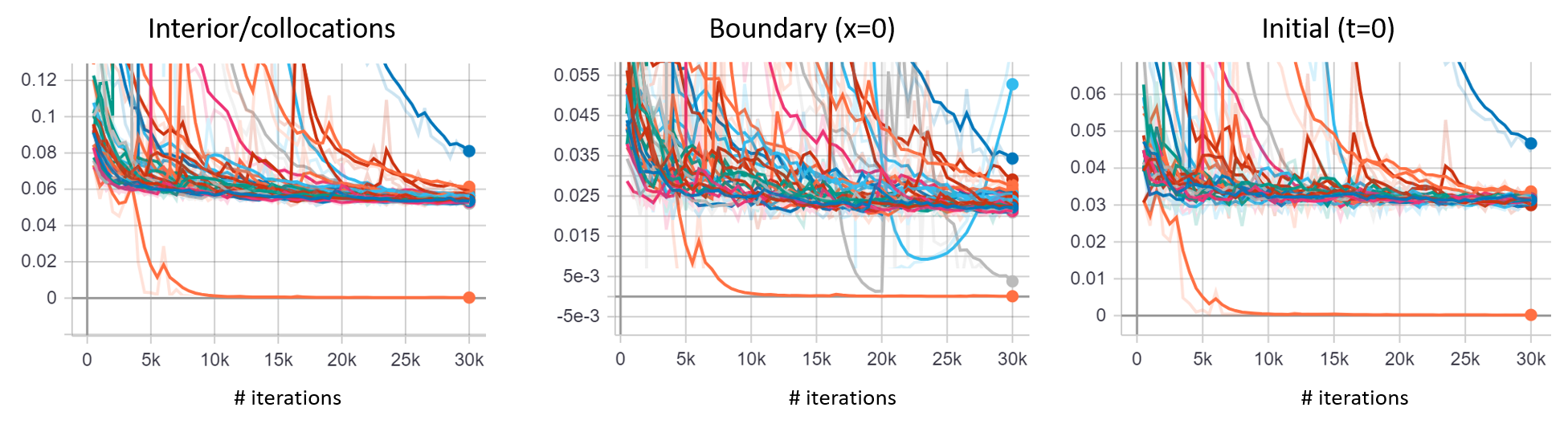}%
    \caption{Evolution of L2-loss with number of iteration for: collocation points (left), boundary condition (x=0) and initial condition (t=0). All curves except for the bottom (orange) one correspond to experiments with the network architecture. The orange one correspond to the addition of entropy condition to the training}%
    \label{fig:BL_loss_comparison}%
\end{figure}
The loss function comparison shows that all the advanced deep learning architectures tested did meet some limitations in their ability to reduce training loss. We can see the formation of a lower "plateau" for all the loss functions but one.
On the other hand, we observe a clear improvement when the entropy condition is added to the problem. This result indicates that substantial improvements in the applicability of PINNs can be achieved through physical reasoning rather than through pure network parameters tuning.

\section{Non Uniform initial condition}
The Method of Characteristics (MOC) can be applied in a setting where the initial condition is uniform. Only then are we able to parameterize the convex hull of the flux function presented in figure~\ref{fig:frac_flow_welge}. In the absence of such condition, current methods of resolution involve a discretization of the space time domain and the application of a proper finite volume scheme. For such a hyperbolic equation (Riemann problem), the Godunov scheme (\cite{Godunov1959}) is considered to be the most stable as it does honor the entropy condition and solves the Riemann problem exactly for each grid block. It is non diffusive, total variation diminishing and monotone preserving. 

\subsection{Numerical methods to improve the convergence of PINNS}
Although network architecture choices presented in the previous section do not impact convergence substantially, there are "tricks" one can use to help improving the training along with the accuracy of the solution obtained.
\subsubsection{LBFGS Optimizer}
\begin{figure}%
    \centering
    \includegraphics[width=1\linewidth]{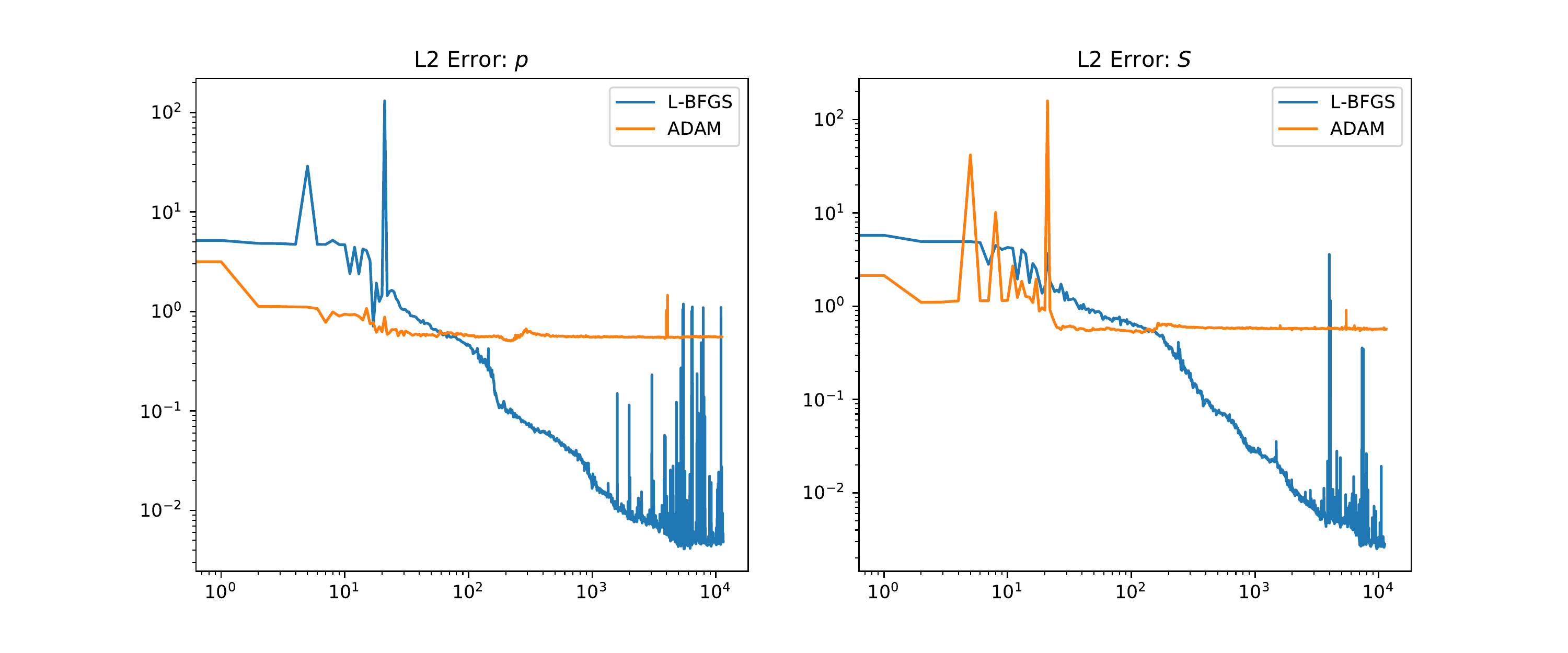}%
    \caption{Evolution of L2-loss with number of iteration for: pressure (left), saturation (right) for ADAM and L-BFGS optimizers}%
    \label{fig:LBFGS_ADAM_loss_comparison}%
\end{figure}

The standard for training neural network is to minimize the loss function using a replacement optimization algorithm for stochastic gradient descent the ADAM optimizer (\cite{Adam2017}). It can handle sparse gradients on noisy problems, features an adaptive stepping strategy and implements a momentum for smoothing. The Broyden-Fletcher-Goldfarb-Shanno (BFGS) algorithm is a class of quasi-Newton methods that uses the Hessian information (2nd derivative) to accelerate the minimization. The L(imited)-BFGS (\cite{ByrdNocedal1995}, \cite{Nocedal1997}) refines at each step an approximation of the Hessian and its principal eigen-components. 
We compared the convergence of both ADAM and L-BFGS for a data rich problem with the interpolation of a pressure and saturation plume on a 2D setup with 3000 gridblocks and 800 timesteps (~5 Millions degrees of freedom). The L2 error for both optimizers are plotted in figure~\ref{fig:LBFGS_ADAM_loss_comparison}

\begin{figure}%
    \centering
    \includegraphics[width=1\linewidth]{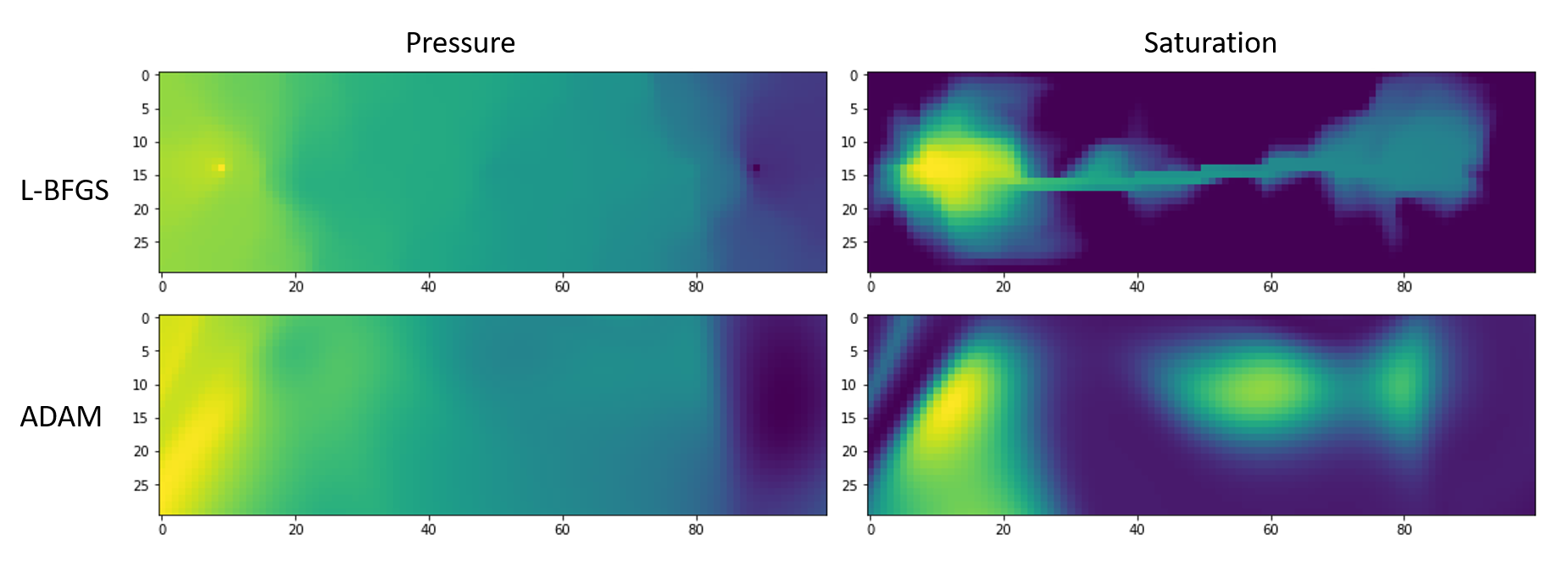}%
    \caption{Pressure and saturation maps at t=607 days for a channelized reservoir. The results are obtained using the ADAM (bottom) and L-BFGS (top) optimizer}%
    \label{fig:LBFGS_ADAM_map_comparison}%
\end{figure}

The error plots show that through ADAM reaches a plateau quite rapidly, the L-BFGS optimizer is capable of finding a lower minimum. The resulting distribution plots are presented in figure~\ref{fig:LBFGS_ADAM_map_comparison}

Although L-BFGS does perform better for smaller scale problems, it may suffer for larger scale problems as it does not implement mini-batching as easily as other stochastic gradient descent algorithms.  

\subsubsection{Weighting of losses}
The key of the PINN approach is to weigh in various objectives (boundary, initial, interior) with associated constraints. The weighting of these losses is a key component to successfully solve partial differential equations. As an example, the losses involved in the Buckley Leverett problem are the following:
\begin{equation}
    \label{eq:Loss_overall}
    \mathcal{L} = \lambda_{res}\mathcal{L}_{res} + \lambda_{BC} \mathcal{L}_{BC} + \lambda_{init} \mathcal{L}_{init}
\end{equation}
where the subscripts $res$, $BC$, $init$ correspond to residual, boundary condition and initial condition. 
This formulation of the loss can be seen as the application of the Lagrange multipliers method in a constrained optimization problem. We can vary the weights between the various components of the loss to ensure that they are all minimized by the optimizer. A very simple rule is to weight the loss inversely to the number of sample used to calculate it: $\lambda_{k} = \frac{1}{N_k}$

We can also apply weights within the sum in the time space domain. Empirically, it appears that lowering the weighting on sharp gradient helps with training. These typically occur at boundaries or in the case of Buckley-Leverett, at the shock level. This method helps with the resolution of the non uniform initial condition problem. The most common weighting used is the signed distance function (SDF) of the geometries featuring sharp edges and corners.
In the forward Riemann problem (Eq.~\ref{eq:residual_buckley}, \ref{eq:frac_flow}, \ref{eq:BL_uniform_bc}), the various losses can be written as:
\begin{equation}
    \begin{aligned}
        \mathcal{L}_{res} &= \sum\limits_{i=1}^{N_{res}}\omega_{res}(x^{(i)}, t^{(i)})\left(\frac{\partial S_{\theta}(x^{(i)}, t^{(i)})}{\partial t} + \frac{\partial f\left(S_{\theta}(x^{(i)}, t^{(i)})\right)}{\partial x} \right)^2\\
        \mathcal{L}_{BC} &= \sum\limits_{i=1}^{N_{bc}}\omega_{bc}(x^{(i)}, t^{(i)})\left(S_{\theta}(x^{(i)}=0, t^{(i)}) - S(x^{(i)}=0, t^{(i)})\right)^2\\
        \mathcal{L}_{init} &= \sum\limits_{i=1}^{N_{init}}\omega_{init}(x^{(i)}, t^{(i)})\left(S_{\theta}(x^{(i)}, t^{(i)}=0) - S(x^{(i)}, t^{(i)}=0)\right)^2\\
    \end{aligned}
\end{equation}

\subsubsection{Integral Continuity planes}
In the absence of loss or source terms, fluid flow problems must honor conservation of mass. It is not given in the PINNs formulation and it can be useful to specify it as a constraint to the problem. One can specify such conservation by making sure that the mass (or volume in incompressible flow) through some "continuity planes" within the domain remains the same. This can be achieved by measuring statistical integral of the quantity of interest and specifying as a constraint that the mass imbalance be minimized.

\subsection{Problem}
We solve Eq.~\ref{eq:residual_buckley} subject to non-uniform initial conditions:
\begin{eqnarray}
    S(x=0,t) &= 0\\
    S(x,t=0) &= x
\end{eqnarray}

As it is not possible to define a welge for the flux function, various parametrization techniques were implemented to solve this problem. We add an artificial diffusion term to the equation~\ref{eq:residual_buckley} first. It becomes:
\begin{equation}
\label{eq:residual_buckley_diffusion}
    \frac{\partial S}{\partial t} + \frac{\partial f(S)}{\partial x} = \epsilon\frac{\partial^2 S}{\partial x^2}
\end{equation}
\cite{Fuks2020} presented a comparative study of this for the uniform initial boundary. It shows that the added diffusion helps with convergence (as it enforces the uniqueness of the solution) and that the multiplying factor $\epsilon$ has a strong influence on the behavior of the objective function. This can be seen as an additional weighting applied to a new constraint and added to the general loss function we seek to minimize. The disadvantage of such an approach is that it introduces a slight smearing at the shock. We resort to this by introducing a parameterization of the diffusion factor. We use a decreasing exponential factor starting with values between $10^{-2}$ and $10^{-3}$ and let the factor decrease geometrically after each iteration.
We find that the diffusion is helpful in the earlier part of the optimization. Hence, we use a vanishing diffusion that tends towards zero. This leads to a solution of the pure hyperbolic problem. This is not an entirely satisfactory solution as it still requires a manual calibration.

\begin{figure}%
    \centering
    \includegraphics[width=1\linewidth]{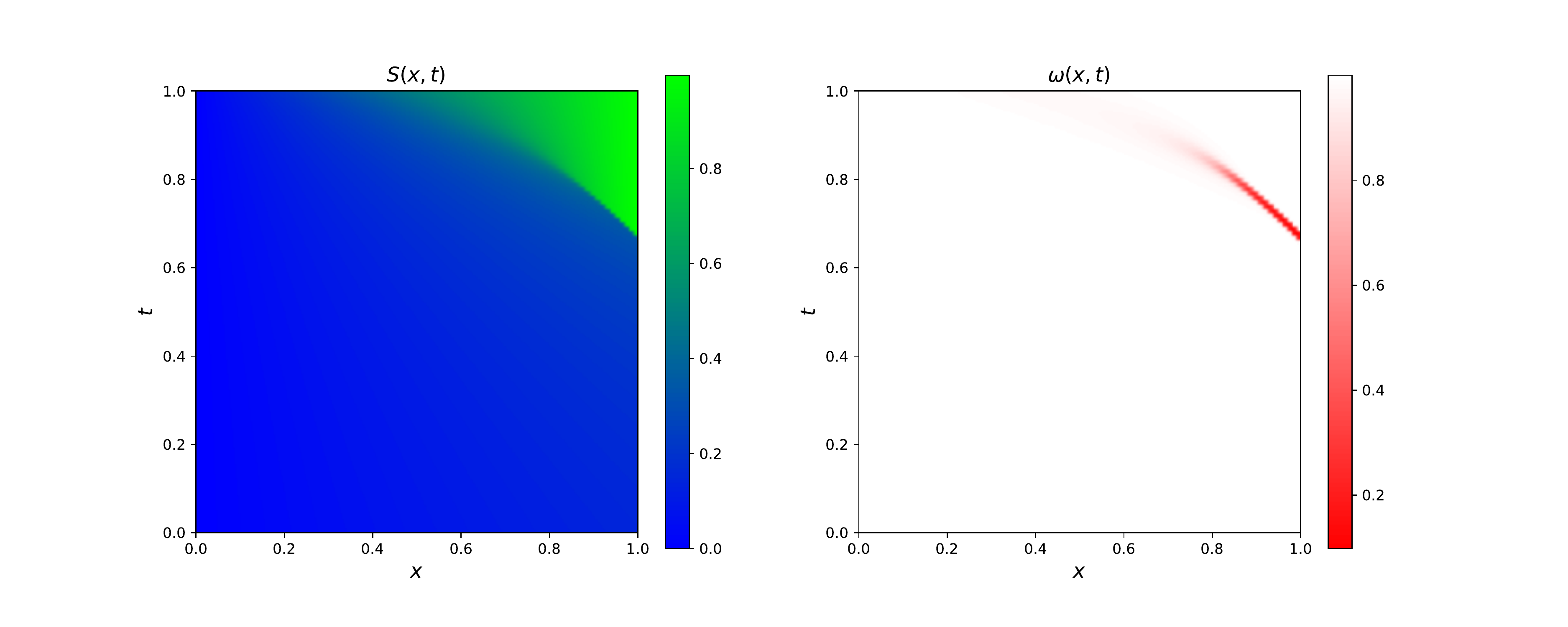}%
    \caption{Distribution of the saturation (left) and associated equation weights (right) in the $(t,x)$ domain. Lower weights are attributed to sample located around sharp gradients}%
    \label{fig:weights_equation}%
\end{figure}

\begin{figure}%
    \centering
    \includegraphics[width=1\linewidth]{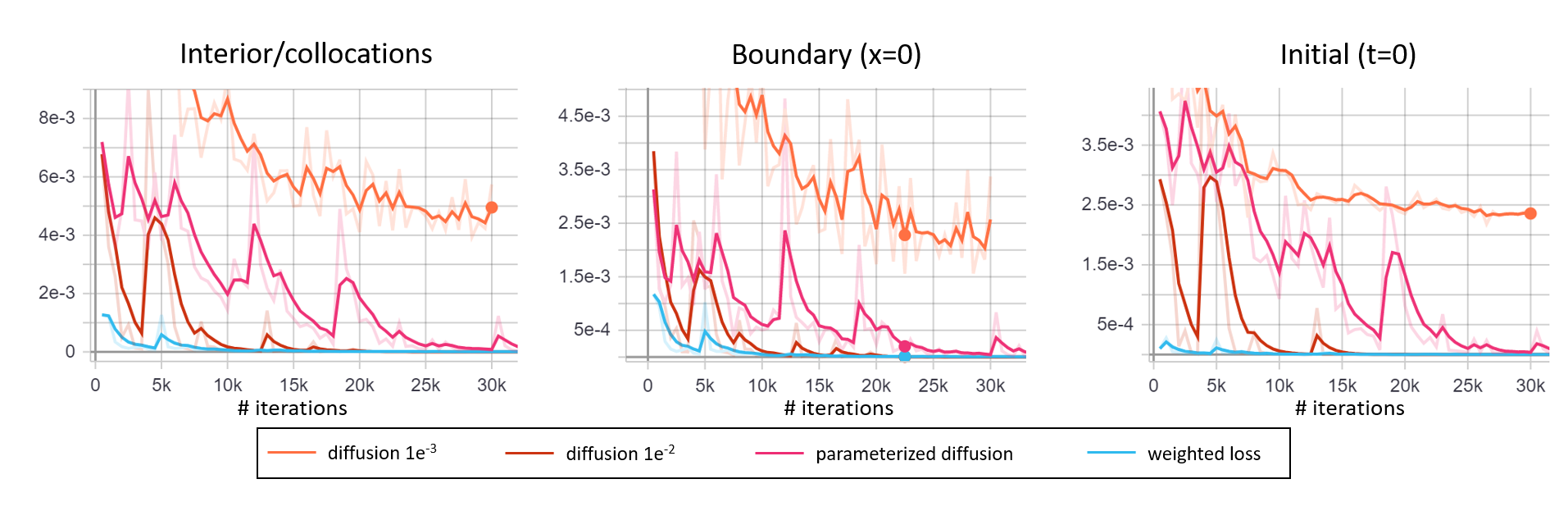}%
    \caption{Evolution of L2-loss with number of iteration for: collocation points (left), boundary condition (x=0) and initial condition (t=0) for Buckley Leverett Problem with non uniform initial conditions}%
    \label{fig:BL_nonuniform_loss_comparisons}%
\end{figure}

A spatial weighting of the loss (SWL) function was then introduced. We weight the equations inversely to the magnitude of the gradients. This is accomplished by multiplying the loss from the residual equation with a factor $\omega(x,t)$ given by:
\begin{equation}
    \omega(x,t) = \frac{1} {(\partial S/\partial x)^2 + (\partial S/\partial t)^2 + 1}
\end{equation}
This effectively lowers the weight of the loss terms nearby the shock wave and helps with convergence. The weight contribution on the domain is represented in figure \ref{fig:weights_equation}.
The residual loss becomes:
\begin{equation}
    \mathcal{L}_{res} = \sum\limits_{i=1}^{N_{res}}\omega(x^{(i)}, t^{(i)})\left(\frac{\partial S_{\theta}(x^{(i)}, t^{(i)})}{\partial t} + \frac{\partial f\left(S_{\theta}(x^{(i)}, t^{(i)})\right)}{\partial x} \right)^2
\end{equation}
The evolution of the various losses (residual, boundary, initial) during training are represented in figure~\ref{fig:BL_nonuniform_loss_comparisons}. 

\begin{figure}%
    \centering
    \includegraphics[width=1\linewidth]{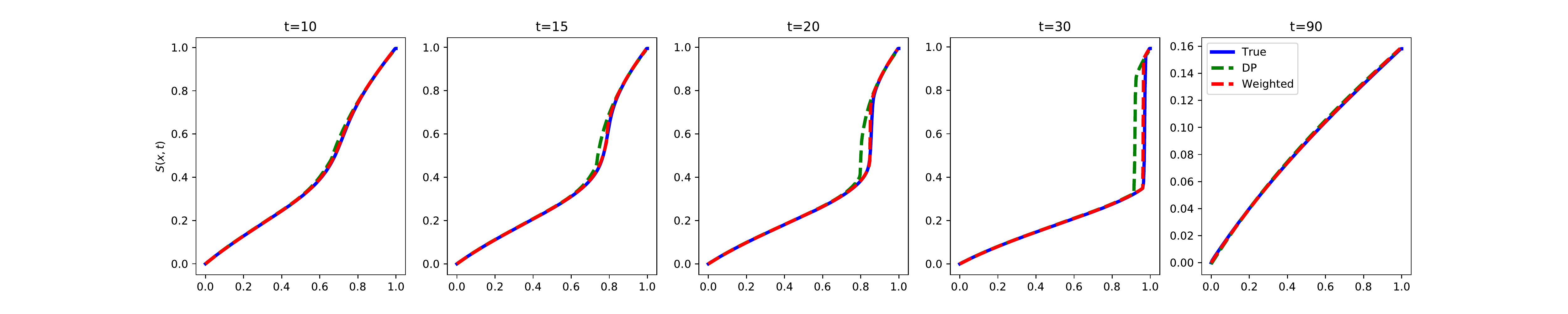}%
    \caption{Results on inference case with non uniform initial condition. The true solution is plotted in blue while the weighted loss is in red and the parameterized diffusion (DP) in green.}%
    \label{fig:BL_Linear}%
\end{figure}

We compare the solutions obtained using a finite volume solver employing a Godunov scheme, the parameterized diffusion and weighted loss solutions in figure~\ref{fig:BL_Linear}. The solution with SWL leads to the smallest overall error, shows a better fit with the finite volume reference and a more accurate capture of the shock.

\section{Gravity Segregation}
We extend the displacement formulation to a flow problem with gravity segregation. We present cases in two dimensions. We consider a two phase system with immiscible water and $CO_2$ for a carbon sequestration in a saline aquifer.
Tchelepi et. al (\cite{ERE221}) provides a comprehensive overview of the presented problem. The Darcy velocity in the presence of gravity can be formulated for each of the phases separately. We impose Neuman boundary conditions with no flow in the horizontal direction. This leads to the vertical velocities, $v_{c,w}$, of $CO_2$ and water from Darcy's law and relative permeability as:
\begin{equation}
\begin{aligned}
    v_w &= -\frac{kk_{rw}}{\mu_w}\left(\frac{\partial p_w}{\partial z} + \rho_wg\right),\\
    v_c &= -\frac{kk_{rc}}{\mu_w}\left(\frac{\partial p_c}{\partial z} + \rho_cg\right)\\
    \label{eq:frac_velocities}%
\end{aligned}
\end{equation}

Mass conservation:
\begin{equation}
\begin{aligned}
    \phi\frac{\partial S_w}{\partial t} + \frac{\partial v_w}{\partial z} &= 0,\\
    \phi\frac{\partial S_c}{\partial t} + \frac{\partial v_c}{\partial z} &= 0\\
    S_w + S_c &= 0\\
    \label{eq:mass_conservation}
\end{aligned}
\end{equation}

If we combine eq.\ref{eq:mass_conservation} and eq.\ref{eq:frac_velocities}, we obtain
\begin{equation}
    \phi\frac{\partial (S_w + S_c)}{\partial t} + \frac{\partial (v_w + v_c)}{\partial z} = 0
\end{equation}

Since $S_w +S_c = 1$, we obtain the uniformity of the total fractional velocity $v_w+v_c = const$. Since our boundary conditions imposed at the top and bottom of the domain are no flow, we collect that

\begin{equation}
\begin{aligned}
    v_c + v_w &= 0,\\
    v_c &= -v_w\\
    \label{eq:counter_current_flow}
\end{aligned}
\end{equation}
This conditions is also known as counter-current flow and impose that in a buoyant system with two phases, the fractional velocities of each phase are opposed.

We can re-arrange eq.~\ref{eq:frac_velocities}:

\begin{equation}
\begin{aligned}
    -\frac{\partial p_w}{\partial z} &= -\frac{\mu_wv_w}{kk_{rw}} + \rho_wg,\\
    -\frac{\partial p_c}{\partial z} &= -\frac{\mu_cv_c}{kk_{rc}} + \rho_cg,\\
    \label{eq:frac_velocities_rearg}%
\end{aligned}
\end{equation}

If we ignore capillary effect, eq.\ref{eq:frac_velocities_rearg} can be further simplified:
\begin{equation}
    \frac{\partial (p_c - p_w)}{\partial z} = \frac{\mu_wv_w}{kk_{rw}} - \frac{\mu_cv_c}{kk_{rc}} + g(\rho_w - \rho_c) = 0
\end{equation}

Using the counter-current flow condition, we obtain:
\begin{equation}
    v_w\left(\frac{\mu_wv_w}{kk_{rw}} + \frac{\mu_cv_c}{kk_{rc}}\right) = g(\rho_c - \rho_w)
\end{equation}

\begin{equation}
\begin{aligned}
    v_w &= \frac{g(\rho_c - \rho_w)}{\frac{\mu_wv_w}{kk_{rw}} + \frac{\mu_cv_c}{kk_{rc}}}\\
    &= \frac{g(\rho_c - \rho_w)k}{\mu_w}\times\frac{k_{rw}}{1+\frac{k_{rw}\mu_c}{k_{rc}\mu_w}}\\
    &= \frac{g(\rho_c - \rho_w)k}{\mu_w}\times f_w(S_w)\\
\end{aligned}
\end{equation}

With a similar reasoning we get the fractional velocity for the $CO_2$ phase:
\begin{equation}
\begin{aligned}
    v_c &= \frac{g(\rho_w - \rho_c)k}{\mu_c}\times\frac{k_{rc}}{1+\frac{k_{rc}\mu_w}{k_{rw}\mu_c}}\\
    &= \frac{g(\rho_w - \rho_c)k}{\mu_c}\times f_c(S_c)\\
\end{aligned}
\end{equation}

We can now replace the fractional velocities in eq.~\ref{eq:mass_conservation}. 

\begin{equation}
\begin{aligned}
    \phi\frac{\partial S_w}{\partial t} + \frac{\partial v_w}{\partial z} &= \phi\frac{\partial S_w}{\partial t} + \frac{\partial}{\partial z}\left(\frac{g(\rho_c - \rho_w)k}{\mu_w}  \frac{k_{rw}}{1+\frac{k_{rw}\mu_c}{k_{rc}\mu_w}}  \right)&= 0\\
\end{aligned}
\end{equation}
We introduce the fractional flow: 
\begin{equation}
    \begin{aligned}
        f_w &= \frac{v_w}{v_c + v_w}\\
        &= \frac{k_{rw}}{1+\frac{k_{rw}\mu_c}{k_{rc}\mu_w}}\\
    \end{aligned}
\end{equation}

\begin{equation}
    \phi\frac{\partial S_w}{\partial t} + \frac{\partial}{\partial z}\left(\frac{g(\rho_c - \rho_w)k}{\mu_w} f_w(S_w)\right)= 0
\end{equation}

If we consider the fluids incompressible (constant $\rho$) and the field properties uniform (constant $k$ and $\phi$), we can separate the fractional flow from the gravity term.

\begin{equation}
\begin{aligned}
    \phi\frac{\partial S_w}{\partial t} + \frac{\partial v_w}{\partial z} &= \phi\frac{\partial S_w}{\partial t} + \frac{g(\rho_c - \rho_w)k}{\mu_w} \frac{\partial f_w(S_w)}{\partial z}  &= 0\\
    & \frac{\partial S_w}{\partial t} + \frac{g(\rho_c - \rho_w)k}{\phi\mu_w} f_w^\prime(S_w)\frac{\partial S_w}{\partial z}  &= 0 \\
\end{aligned}
\end{equation}

Similarly for the $CO_2$ phase:
\begin{equation}
    \frac{\partial S_c}{\partial t} + \frac{g(\rho_w - \rho_c)k}{\phi\mu_c} f_c^\prime(S_c)\frac{\partial S_c}{\partial z}  = 0
\end{equation}

In the horizontal direction, note that we recover the Buckley Leverett flow written in equation \ref{eq:residual_buckley}:
\begin{equation}
\begin{aligned}
    \frac{\partial S_w}{\partial t} + f_w^\prime(S_w)\frac{\partial S_w}{\partial x}  &= 0 \\
    \frac{\partial S_c}{\partial t} + f_c^\prime(S_c)\frac{\partial S_c}{\partial x}  &= 0 \\
\end{aligned}
\end{equation}

This allows us to formulate a two dimensional residual for the loss function we are minimizing.

\begin{figure}%
    \centering
    \includegraphics[width=.9\linewidth]{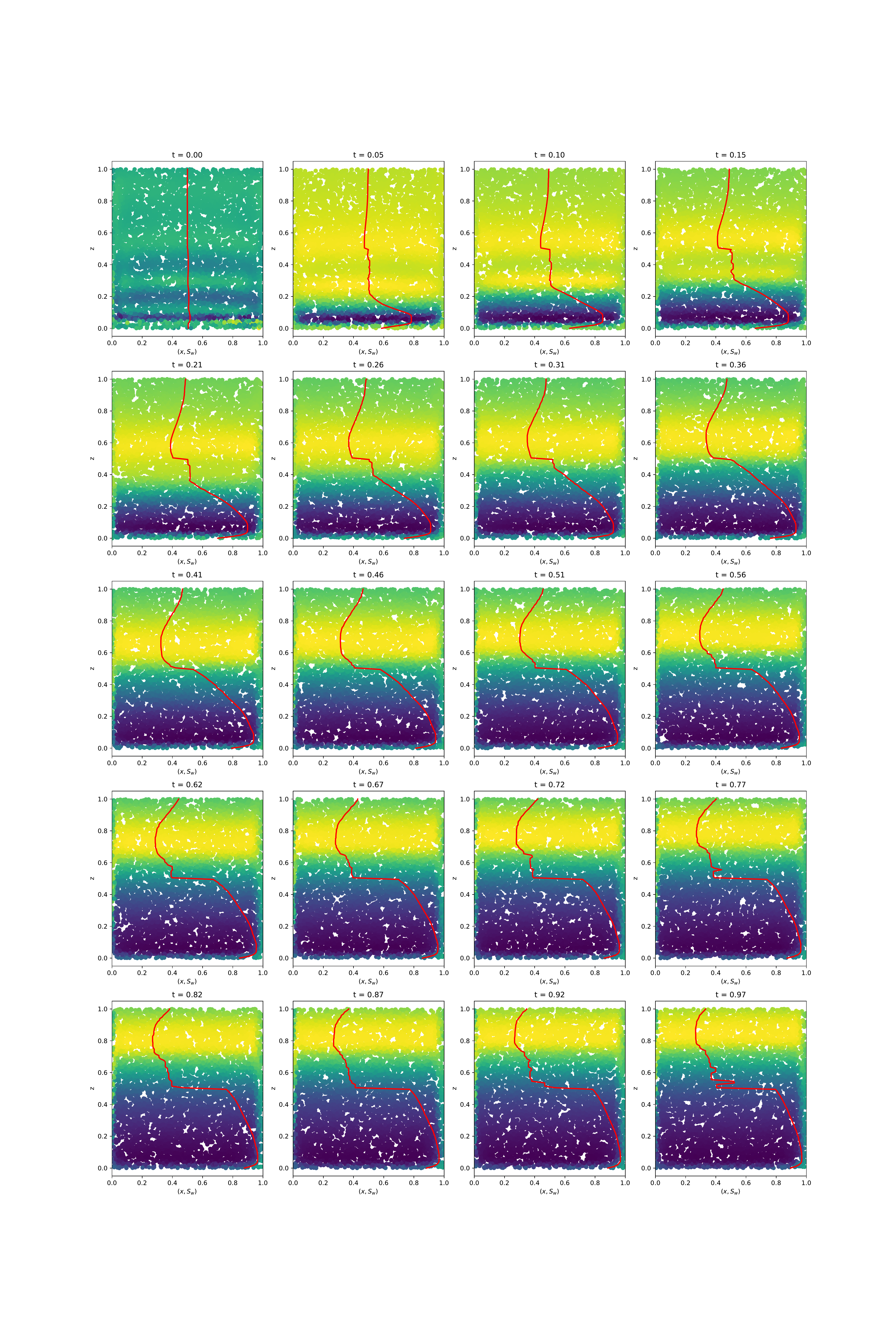}%
    \caption{Evolution of the saturation of two phases through time during the process of segregation along with a 1D projection of the average saturation (red) along the $z$ axis}%
    \label{fig:2d_uniform_seg}%
\end{figure}

The first application case is a vertical segregation where the initial condition is a uniform mixed saturation for the water and $CO_2$ phase of 0.5. Figure \ref{fig:2d_uniform_seg} shows the phases distributions during the process of the segregation.

If the general behavior of the simulation is correct, the solution displays local inconsistencies such as oscillations and non-monotonousness that are not physical. A one dimensional projection of the average behavior along the $z$ axis highlights such inconsistencies. This can be attributed partly to an inaccurate treatment of the boundary conditions. As explained earlier, the optimization process does not perform well on sharp gradients or areas of discontinuity such as close boundaries. For this reason, we implement a loss function that emulates an open boundary condition and make the permeability operator constant everywhere in the domain except at the boundaries where it tends towards zero. This practically imposes a near no-flow at the boundary while avoiding the treatment of such discontinuities using spatial weighting of the losses. This permeability operator can also be learnt as shown next.

\subsection{Non-uniform permeability operator}
We are interested in emulating a non uniform permeability operator using a deep network. Indeed, the interpolation capabilities of network make them an ideal tool to produce a functional form $k_{\theta}(x,z)$ that is both continuous, differentiable and closely fits a heterogeneous permeability field $\textbf{k}$ in high dimension ( $d_x\times d_z >> 1$ for the two dimensional case considered).

\begin{align*}
  \mathcal{N} \colon \mathbb{R}^{d_x\times d_z} &\to \mathcal{C}^\infty(\mathbb{R}^2)\\
  \textbf{k} &\mapsto k_{\theta}(x,z).
\end{align*}

This alleviates the need for a discretization in heterogeneous cases. Indeed, the mass conservation equation in eq.\ref{eq:mass_conservation} can be written:
\begin{equation}
\begin{aligned}
    \phi\frac{\partial S_w}{\partial t} - \frac{\partial}{\partial x}\left(\frac{k_{\theta}(x,z)k_{rw}}{\mu_w}\frac{\partial p_w}{\partial x} \right) - \frac{\partial}{\partial z}\left[\frac{k_{\theta}(x,z)k_{rw}}{\mu_w}\left(\frac{\partial p_w}{\partial z} + \rho_wg\right)\right]  &= 0\\
    \phi\frac{\partial S_c}{\partial t} - \frac{\partial}{\partial x}\left(\frac{k_{\theta}(x,z)k_{rc}}{\mu_c}\frac{\partial p_c}{\partial x} \right) - \frac{\partial}{\partial z}\left[\frac{k_{\theta}(x,z)k_{rc}}{\mu_c}\left(\frac{\partial p_c}{\partial z} + \rho_cg\right)\right]  &= 0
\end{aligned}
\label{eq:mass_conservation_2D}
\end{equation}
with the permeability $k_{\theta}(x,z)$ fully defined as a composition of linear/non-linear transformations typically formed in a multi-layer network. Such an expression can be:
\begin{equation}
    k_{\theta}(x,z) = tanh\left( \mathbf{W}^{[2]} \left[ tanh \left( \mathbf{W}^{[1]} \left[ tanh \left( \mathbf{W}^{[0]} [x,z]^T + b^{[0]} \right) \right] + b^{[1]} \right) \right] + b^{[2]} \right)
    \label{eq:perm_MLP}
\end{equation}

\begin{figure}%
    \centering
    \includegraphics[width=1\linewidth]{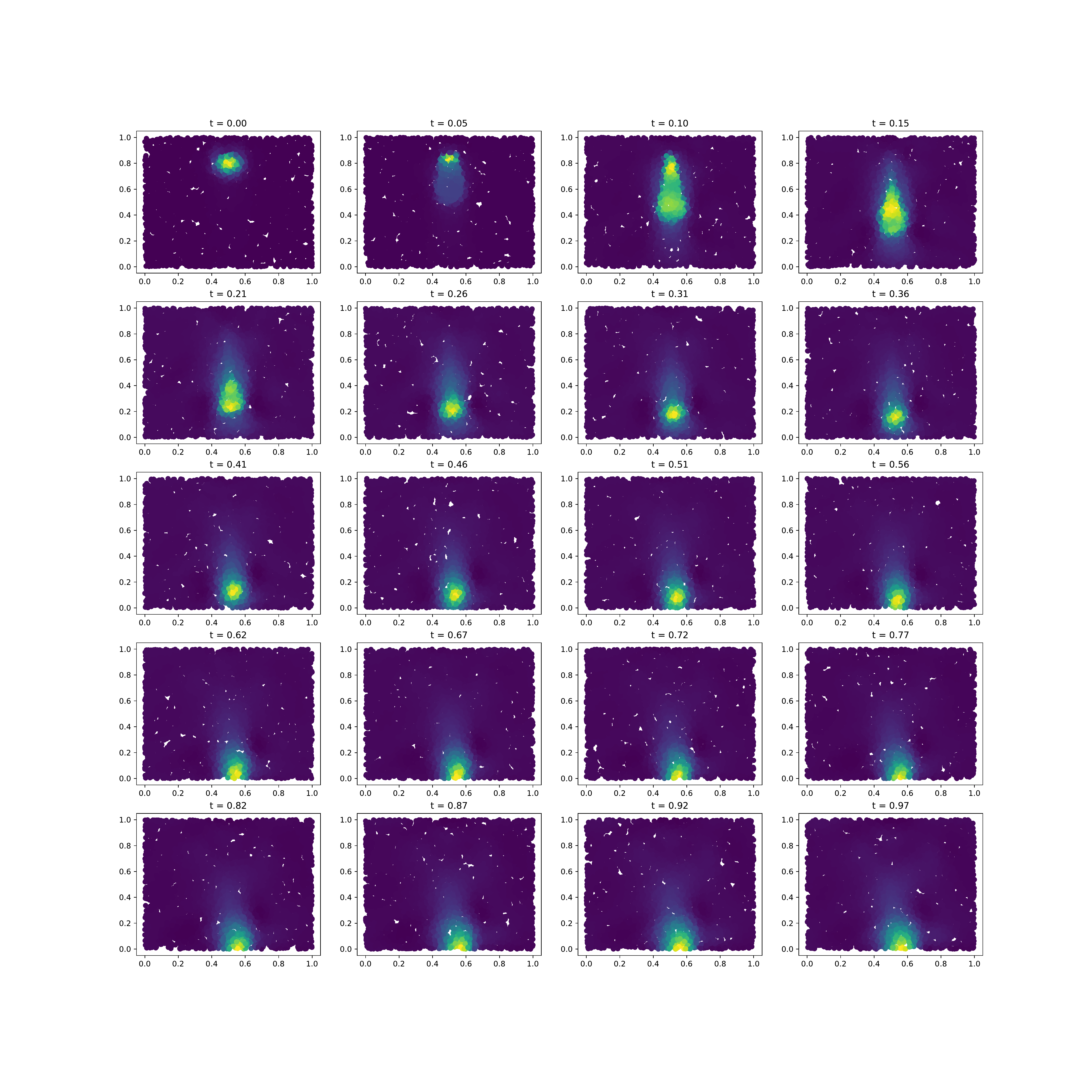}%
    \caption{Evolution of the saturation of a drop of fluid into a less dense phase using PINNS.}%
    \label{fig:2d_drop}%
\end{figure}

The network used to emulate the permeability field can be trained in a pre-processed step or jointly with the flow and transport conservation equation. We illustrate this example with a "drop fall" example. Figure \ref{fig:2d_drop} shows the evolution of the saturation field at various time steps.

\begin{figure}%
    \centering
    \includegraphics[width=1\linewidth]{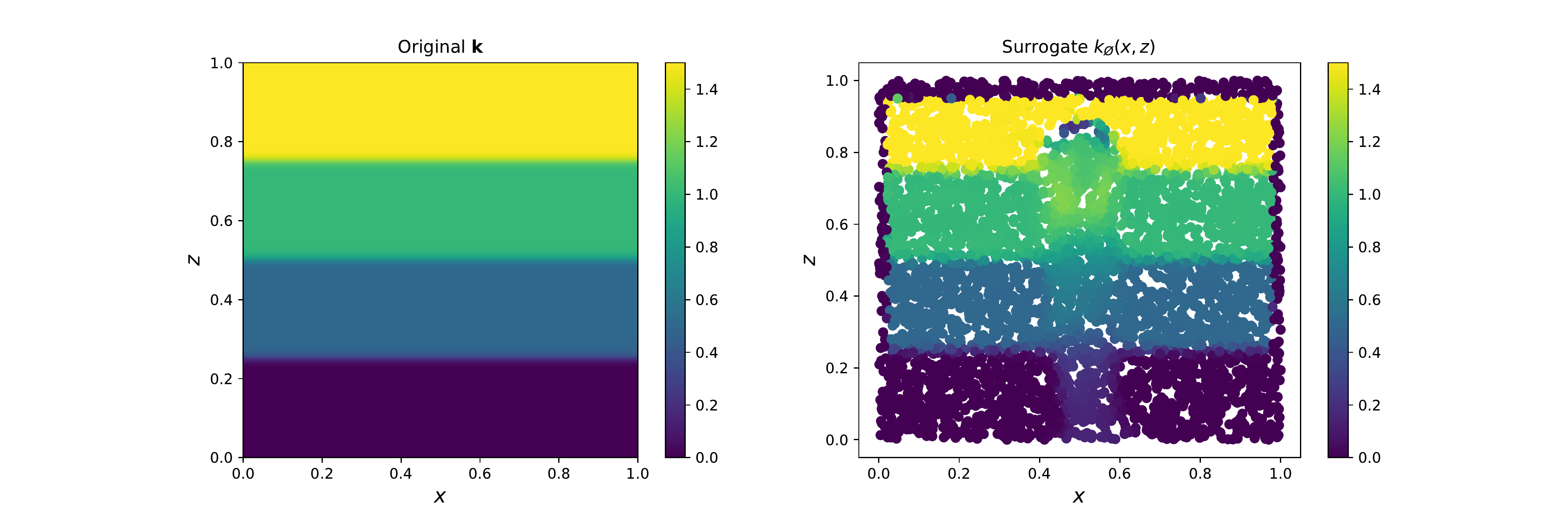}%
    \caption{Original permeability field multiplier (left) and surrogate obtained with a neural network (right)}%
    \label{fig:layered_perm}%
\end{figure}

The permeability field used for this example is composed of various layers of decreasing values. Figure\ref{fig:layered_perm} shows the original permeability field along with the one approximated using a neural network. This network is trained jointly with the simulation but not used in the back-propagation as the gradient computed for it does not propagate backward. This guarantees that we do not minimize the loss function based on the permeability field. If we were in an inversion setup, we could easily switch to a back-propagating permeability which could be conditioned by additional data.

One implementation consideration is to make sure that during the resolution of the PDE, back-propagation does not pass through the permeability operator. This causes the approximated permeability field to present artefacts of the solution we simulate. This is illustrated in figure~\ref{fig:layered_perm} where we see the trail of the drop forming on the permeability field while it is supposed to be static.

\subsection{Performance and Error}
\begin{figure}%
    \centering
    \includegraphics[width=1\linewidth]{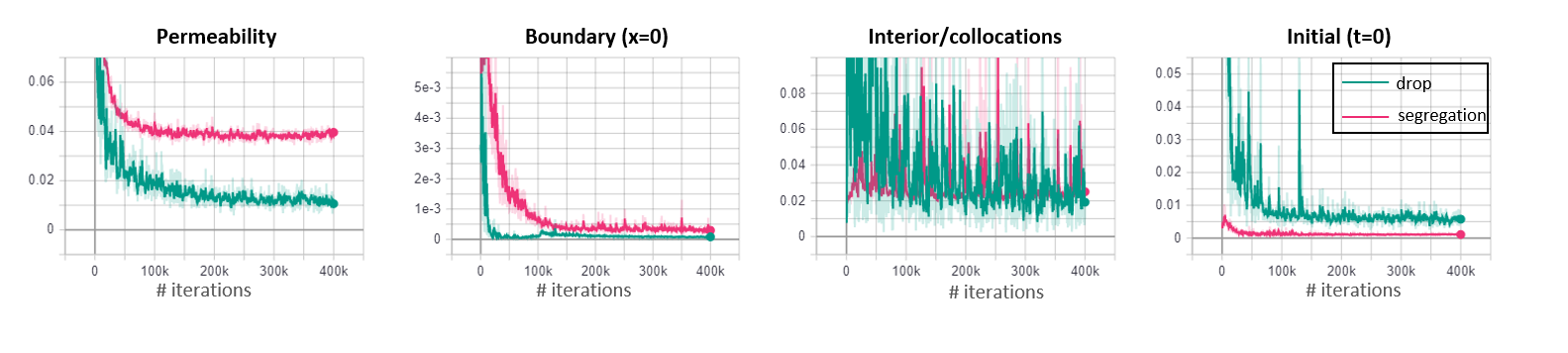}%
    \caption{Evolution of $L_2$-loss with number of iteration for (from left to right): permeability operator, boundary condition (x=0), collocation points and initial condition (t=0) for two dimensional gravity segregation problem with uniform initial conditions (pink) and single drop (green)}%
    \label{fig:loss_comp_2d}%
\end{figure}

We design the simulation in a way that maximizes memory utilization on a given hardware. This guarantees that the number of sampled points is maximized for a given run. We use a Tesla V100 GPU with 16 Gigabytes of memory. In one or two dimensional cases, we do not need to resort to mini-batch gradient descent as the sample size can be decreased without affecting the overall accuracy of a run. Figure~\ref{fig:loss_comp_2d} shows the evolution of the various losses functions with number of iterations.

We note that the two dimensional runs take significantly more iterations to converge (about 400,000 in the case presented). The average run time is about 30ms per iteration. This results in a total run time of approximately 3 hours. Although this is much slower than a finite volume simulation, the advantage is that every iteration is performed on the entire space time domain at once. This means that we can start using the model and informing decisions before the simulation has ended.

\section{Conclusion and Future Work}
We have demonstrated that a Physics Informed Deep Learning approach can be utilized to solve non linear hyperbolic partial differential equations with non convex flux terms (Riemann problem). We show that a rigorous definition of the problem's boundary and entropy conditions is necessary for the proper integration of the PDE using this new class of numerical methods. We presented the results of extensive tests using various network's architectures such as Fourier Network, DGM and Periodic Networks (SiRen) and showed that they did not bring significant improvement for the problem of interest. We showed on the other end that a series of techniques involving the implementation of proper constraints as additional loss functions complemented by a proper weighting does result in significant gains in terms of training. 

We illustrate our findings with a series of physical experiments in one and two dimensions. If the results in one dimension establish a new state of the art, improvements are expected for the two dimensional cases. We believe that they are related to the same issues encountered in one dimension and are related to the existence of sharp discontinuities inherent to this class of hyperbolic problems.

Our future work will focus on the extension of the current approach to two and three dimensions, the proper handling of boundary conditions and the parameterization of heterogeneities using the grid-less approach demonstrated in this work.

\section{Acknowledgement}
We thank Mr Adrien Papaioannou for providing useful code review for the implementation of the experimental PINN, Oliver Hennigh and Sanjay Choudhry from NVIDIA for the introduction to SimNet and the the help implementing a Riemann problem with entropy constrained flux function.